# A Novel Method for Detecting Dust Accumulation in Photovoltaic Systems: Evaluating Visible Sunlight Obstruction in Different Dust Levels and AI-based Bird Droppings Detection

Md Shahriar Kabir,[1, a)] Khalid Mahmud Niloy,[1] S. M. Imrat Rahman,[2] Md Imon Hossen,[1], Sumaiya Afrose,[1] Md. Ismail Hossain Mofazzol,[3] and Md Lion Ahmmed[4]

[1] Electrical and Electronics Engineering (EEE), American International University Bangladesh (AIUB), Dhaka, Bangladesh;

[2] Holcombe Department of Electrical and Computer Engineering, Clemson University, Clemson, SC 29634, USA;

[3] Computer Science Engineering (CSE), Uttara University, Dhaka, Bangladesh;

[4] Electrical and Electronics Engineering (EEE), European University of Bangladesh, Dhaka, Bangladesh;

[a)] **Author to whom correspondence should be addressed:** tulip6585@gmail.com

# A Novel Method for Detecting Dust Accumulation in Photovoltaic Systems: Evaluating Visible Sunlight Obstruction in Different Dust Levels and AI-based Bird Droppings Detection

**Abstract**

This paper presents an innovative method for automatically detecting dust accumulation on a PV system and notifying the user to clean it instantly. The accumulation of dust, bird, or insect droppings on the surface of photovoltaic (PV) panels creates a barrier between the solar energy and the panel's surface to receive sufficient energy to generate electricity. The study investigates the effects of dust on PV panel output and visible sunlight ($V_{SL}$) block amounts to utilize the necessity of cleaning and detection. The amount of blocked visible sunlight while passing through glass due to dust determines the accumulated dust level. Visible sunlight can easily pass through the clean, transparent glass but reflects when something like dust obstructs it. Based on those concepts, a system is designed with a light sensor that is simple, effective, easy to install, hassle-free, and can spread the technology. The study also explores the effectiveness of the detection system developed by using image processing and machine learning algorithms to identify dust levels and bird or insect droppings accurately. The experimental setup in Gazipur, Bangladesh, found that excessive dust can block up to 55% of visible sunlight, wasting 55% of solar energy in the visible spectrum, and cleaning can recover 3% of power weekly. The data from the dust detection system is correlated with the 400W capacity solar panels' naturally lost efficiency data to validate the system. This research measured visible sunlight obstruction and loss due to dust. However, the addition of an infrared radiation sensor can draw the entire scenario of energy loss by doing more research.

**Introduction:**

Sunlight is one of the planet's most plentiful and easily accessible energy sources. The energy from sunlight in a single day equals the planet's overall energy needs for the year. The sun produces almost twice as much energy as traditional sources such as coal, petroleum, and natural and nuclear materials [1]. In recent years, there has been an increase in reliance on solar energy. Countries are considering emphasizing this and developing power plants for that purpose, as well as attempting to spread the usage of this energy in rural regions to address future non-renewable energy deficits. In 2000, the electricity generated from solar energy was

1.3 GW, and the electricity production increased 800% in the next 20 years [2]. As per the International Renewable Energy Agency (IRENA), the total installed capacity of solar power plants worldwide was 714 GW (24.3%) till 2020. At that time, the highest installed capacity of solar plant countries was China (255 GW), the United States (75 GW), Japan (68 GW), Germany (53 GW) and India (39 GW). IRENA fixed a goal by setting a map, "The Remap Case 2050", to achieve an 86% share of renewable energy production within 2050 [3]. In 2021, 150 GW was added to the production [4]. Solar PV and wind power's combined share of installed generation capacity increased by 2.4% in 2022. China reduced the emission intensity of the power sector by 2.5% in 2022 by using renewable energy sources [5]. The predicted electricity generation from PV panels within 2024 is about 1296 GW of solar energy [6]. So, reliance on renewable energy usage is increasing, and it is time to overcome the barrier to improve sustainability.

Dust accumulation decreases the efficiency of PV panels, and the rate becomes extensive if they are not cleaned the panels for a long time. Studies have shown that dust accumulation may reduce solar panel efficiency by 30% to 40%, resulting in a fall in power generation of up to 86% [7]. The short-circuit current, open-circuit voltage, and power production of PV panels all decrease with increasing dust levels [8]. A field test showed that a dust concentration of 10 g/m$^2$ can cause a 38% reduction in PV panels' maximum power [9]. On a photovoltaic panel, dust deposition of 20 g/m$^2$ affects open circuit voltage, short circuit current, and efficiency by 15–21%, 2–6%, and 15–35%, respectively [10]. Under natural exposure conditions, PV performance drops daily, weekly, and monthly by 6.24%, 11.8%, and 18.74% due to the dust [11]. In the US, there was an average of 1%, with a peak of 4.7% during two months of monitoring, a 40% loss in six months in Saudi Arabia, and an 11% reduction in efficiency in Thailand's tropical climate, 29.6% in Kathmandu, Nepal, and a 33.5% to 65.8% drop in efficiency from an Egyptian study [12-15]. However, the dust accumulation is comparatively lower on tilt PV panels than on horizontally fixed panels—about 1%-14% more losses in horizontally fixed panels [16]. So, dust accumulation decreases the PV output, and an alert about high dust accumulation is required to clean the panel.

The production of solar PV is also influenced by the dust's chemical, biological, and electrostatic properties [17]. The dust particles' shape, size, and weight significantly affect how much they accumulate [18]. The dust particle diameter is less than 10 µm, although they vary in length [19]. This dust effect highly depended on geographical location and local environmental conditions. Airborne particle concentration, aerodynamic particle size distribution, site location, and meteorological conditions influence dust accumulation on PV module surfaces [20]. Air pollution is one of the critical environmental factors for dust accumulation. Air pollution mainly contains carbon and bacteria, which are types of dust [21]. Africa is one of the locations with the most significant accumulations of sand and dust on Earth. In contrast, Asia is thought to have among the driest, windiest, and dustiest climates worldwide [22]. Rainfall can often help clean the PV panel. The research found that in dry conditions, days without rain resulted in an average daily efficiency reduction of 0.2%. The PV system's surface requires at least 20 mm of rain to clean it naturally [23]. The meteorological conditions, particularly air quality, are primarily responsible for dust accumulation on photovoltaic surfaces.

The solar spectrum represents the dispersion of electromagnetic energy released by the Sun across various wavelengths. It can be classified into many bands: gamma-ray (10 fm to 1 pm), X-ray (1 pm to 10 nm), ultraviolet (10–380 nm), visible (380–780 nm), infrared (780 nm to 1 mm), microwave (1–15 mm), and radio wavelengths (0.1 mm to 100 m) [24]. However, the three central regions are ultraviolet (UV), visible, and infrared (IR). The distribution of solar energy, which peaked at 5000 Å (500 nm), comprises less than 10% in the ultraviolet, less than 50% in the visible spectrum, and less than 50% in the infrared. The solar energy obtained at the Earth's surface is roughly 1000 W/m² [25]. About 6.8% of the solar energy reaching the surface of Earth is UV light. The remaining 93.2% comprises visible light bands and longer wavelength infrared radiation (IR) [26]. Visible sunlight ($V_{SL}$) makes around half of all the sunshine that reaches the surface of Earth [27]. The most common solar panels are designed to collect sunlight in a particular band of wavelengths of about 850nm. So, it predominantly absorbs visible sunlight ($V_{SL}$), then some IR and UV spectrum for converting to electricity [28].

Some research characterizes various flaws in PV panels, including dust collection. Standard solutions include statistical analysis and linear regression models [29, 30], image processing methods [31, 32], and artificial intelligence approaches such as artificial neural networks [33] and fuzzy logic [34]. Those systems performed well when faults occurred but were not specially designed to detect dust. In another study, an approach is given that combines a deep residual neural network (DRNN) with image processing methods like nonlinear interpolation, equivalent segmentation, and clustering to find particular locations in the PV panel where the dust is being deposited [35].

Yfantis and Fayed proposed a camera system for detecting dust particles in 2014. They have used intelligent cameras equipped with red, green, blue, and infrared capabilities for nocturnal vision. These cameras continually capture images of each panel. The classification system processes the image and determines if the surface requires cleaning in real time. The algorithm used here for detection was the classification vector. This system has shown how to fill up the lack, which is crucial in keeping the solar panel clean [36].

Unluturk et al. (2019) structured an image processing-based assessment of dust accumulation by using the STM32F103C8T6 microcontroller in 2019. In this experiment, three types of artificial dirt were made and dumped on PV, and artificial light was applied. As a result, solar irradiance variations have been observed due to the effect of dirt at different stages. So, it indicates the performance of the PV module by observing voltage and current, i.e., power output. The image processing technique predicted the PV module pollution. For this, a GLCM-based ANN classifier algorithm was used to process PV panel images. The study has played an essential role in detecting dirt on solar panels. However, no solution for bird-dropping detection was found in this research [37].

Klugmann et al. (2020) studied the effects of solar panel pollutants and potential solutions, a topic of significant relevance to renewable energy and solar technology. They simulated the natural and artificial dust deposition on solar panels and measured the resulting dust density per square meter in a laboratory setting. The study revealed that the efficiency of solar panels fluctuates seasonally, with the most significant loss occurring in spring, over 15%. The researchers also examined the dust particles using SEM and EDS analysis, identifying materials that contribute to air pollution. They further demonstrated that artificial dust interacts differently with various glass surfaces of solar panels [38].

Saquib et al. (2020) developed an image-processing system using Artificial Neural Networks for dust detection. This innovative system, which uses a backpropagation algorithm to predict the percentage of dust on the solar panel and the output voltage, shows promising results. The system has a low loss between the actual and predicted power, about 0.000146. Although it has only been tested for 20W solar panels, the potential for broader application is evident [39].

Perez-Anaya et al. (2022) provided an approach that combines statistical analysis and heuristic algorithms for the detection of various dust levels in PV panels. The developed system can detect dust in three levels: no, moderate, and high dust levels. They have gathered data using UV, current, voltage, and temperature sensors to train the model. To establish the intermediate dust accumulation condition in the PV panels, 4 g of dust was distributed on the 1 m$^2$ surface of the PV panel. In comparison, the high dust accumulation condition was attained with a layer of 8 g/m$^2$. The data indicated that 96.5% of the clean PV cases were accurately identified, but the remaining 3.5% were misclassified as moderate dust buildup conditions. In terms of the moderate dust buildup condition, only 94.1% of the tests were accurately recognized, and the missing 5.9% was erroneously labeled as the clean condition. For the high dust deposition condition, 100% of the tests were correctly identified. The detection level is limited, which is the drawback of this system [40].

The existing developed system is able to detect dust on PV panels in specific conditions. Most of the systems are ML-based and trained using either camera-captured images or measured physical and environmental data. The limitation is that the data varied location-wise or weather-wise, which needs to retrain the model. Sometimes, the setup for data taking makes the system complicated. To address those issues, an innovative sensor-based dust detection system is designed, which is simple to reduce complexity and dynamic to adapt to any conditions. The paper aims to develop an auto dust detection system and send notifications to mobile apps to take action on cleaning and recovering power faster. An IoT-enabled light sensor is used under transparent glass, like solar cells, under a 2mm to 4mm thick glass layer [41]. Another light sensor is placed in an open space. Both sensors measured the value of transformed $V_{SL}$ through the glass and calculated the reflected light due to dust. The cleaning process can be initiated based on the amount of dust. This research also showed data on $V_{SL}$ transformation through the glass layer for different dust levels. Bird droppings on PV panels are another reason for the reduction of power output. To detect its presence, an ML model is trained by field images because it is not possible to detect bird droppings with any sensor.

## 2. Materials and Methods

### 2.1 Location and weather of the study site

Field data was collected from Dhaka (lat. 23.88, long. 90.39) and Gazipur (lat. 23.98, long. 90.41) to create the scenario of dust accumulation on PV panels. Both places are situated nearby, and the weather conditions are similar. Fig. 1 and Fig. 3 show the air quality of Dhaka in 2024 to describe how much dust, dirt, soot, smoke, and liquid droplets are carrying the air. The tiny solid or liquid droplets suspended in the air are known as particulate matter (PM). Particulate matter-10 (PM-10) means airborne particles less than 10μm in diameter, and PM-

2.5 means particles less than 2.5 µm in diameter. So, PM-10 is slightly larger and falls more quickly on any surface due to the greater force of gravity. PM-2.5 falling rate is much slower, and the winds are also affected. However, the falling rate will also be high for both cases if the dust or other amount percentage is high in the air. Fig. 2 shows that the average PM-10 level in the air exceeds 500 µg/m³ in January and exceeds 200 µg/m³ in February. On some days, it crosses 600 µg/m3, which is shown in Fig. 1.

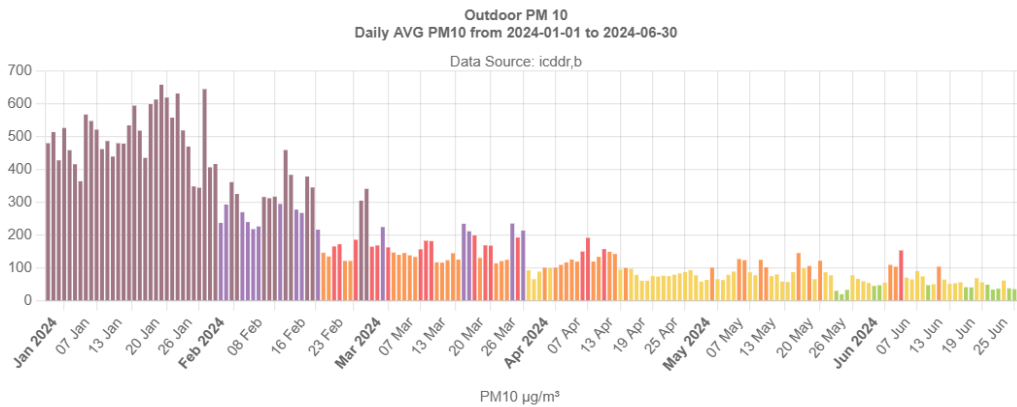

**Fig. 1.** The air quality of Dhaka in 2024 (daily average PM 10) [42].

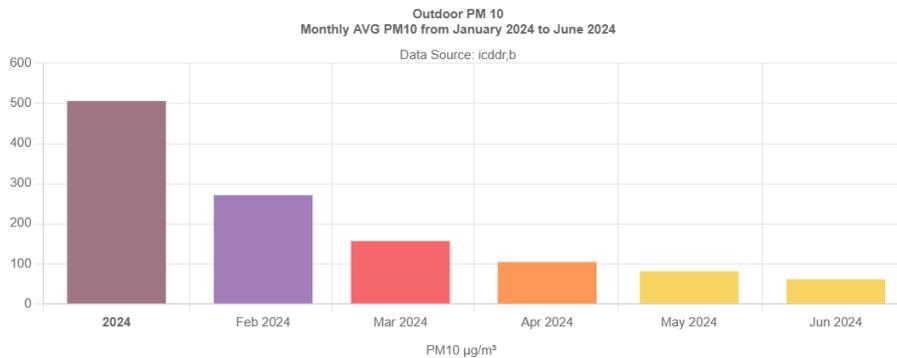

**Fig. 2.** The air quality of Dhaka in 2024 (monthly avg PM 10) [42]

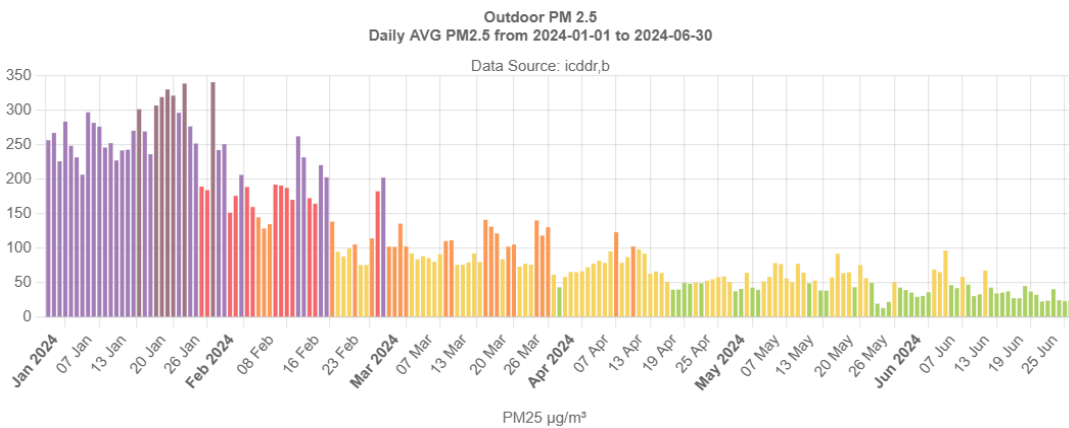

**Fig. 3.** The air quality of Dhaka in 2024 (daily avg PM 2.5) [42]

The March and April months have averaged above 100 µg/m³ PM-10 in the air [42]. So, the higher amount of particulate matter in the air of Dhaka indicated that more dust had fallen on any surface. PM-2.5 is slightly lower than PM-10 but still crosses the danger zone. Fig. 4 shows that January is the top-ranked month for carrying the highest PM-2.5, about 270 µg/m³. After April month, the particulate matter in the air started to decrease due to the rain. So, the air quality remains good during the pre-rainy and rainy seasons in Bangladesh. In Dhaka, rooftop solar panel implementation is a prevalent choice. However, this dust-falling rate reduces the power output, and manual cleaning is the only method. This research is crucial as it highlights the significant impact of dust accumulation on the efficiency of solar panels, thereby emphasizing the need for innovative dust measurement methods in the renewable energy industry.

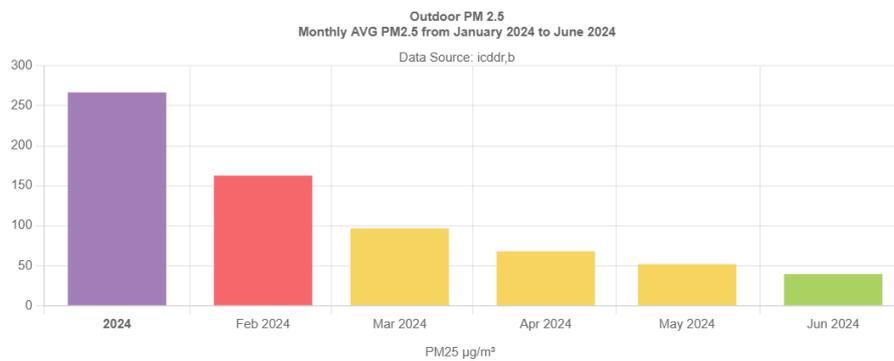

**Fig. 4.** The air quality of Dhaka in 2024 (monthly avg PM 2.5) [42]

Fig. 5 shows Dhaka's average solar radiation and sunshine hours. The average sunshine hour is 7.55, which is enough. June and July have fewer sunshine hours, only around 5. The average solar radiation in Dhaka is 4.73 kWh/m²/day. From Bangladesh's perspective, the average solar radiation in Dhaka is close to Rajshahi, which has the highest average (5 kWh/m2/day) monthly [43].

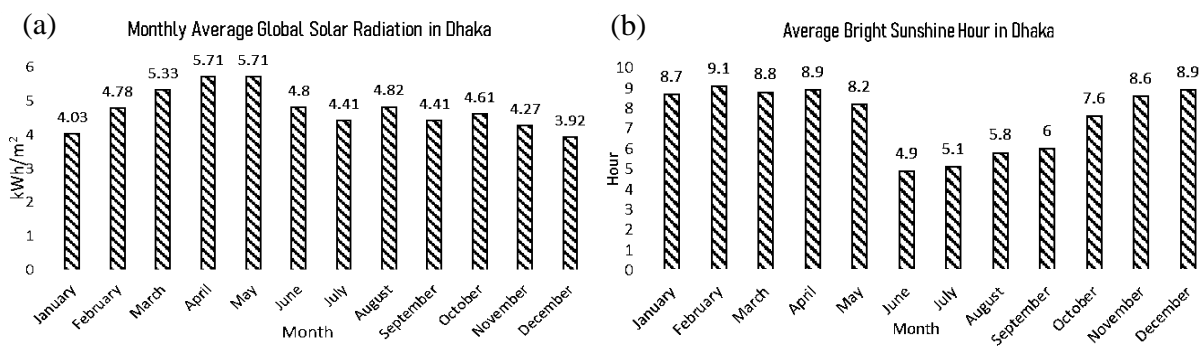

**Fig. 5.** Graphical representation of (a) average solar radiation and (b) average sunshine hour in Dhaka

## 2.2 How Novel Dust Measurement Method Works

Fig. 6 shows a mechanism integral to detecting different dust levels on a surface. This system is constructed innovatively to measure the dust amount, informing the user to initiate the cleaning process. Two light sensors are used here. For this method, a clean, transparent glass is placed beside the solar panel. When dust falls on the transparent glass, the sunlight cannot transform the glass quickly. A light sensor is placed under the transparent glass to measure the $V_{SL}$ passed through the glass. Another light sensor is placed in an open place, controlled by the servo motor to turn ON or OFF. The openly placed light sensor measures direct $V_{SL}$. After measuring both sensor values, they are compared. For high dust on the surface, the open space lux value is higher than the lux value under dusty glass. So, the compared value is high, causing more light to be lost when passing through the glass. For a clean surface, the open space lux value is equal to the lux value under glass. As PV panels have a glass surface above solar cells, this glass method approach replaced solar cells with a light sensor and precisely gave information about dust accumulation.

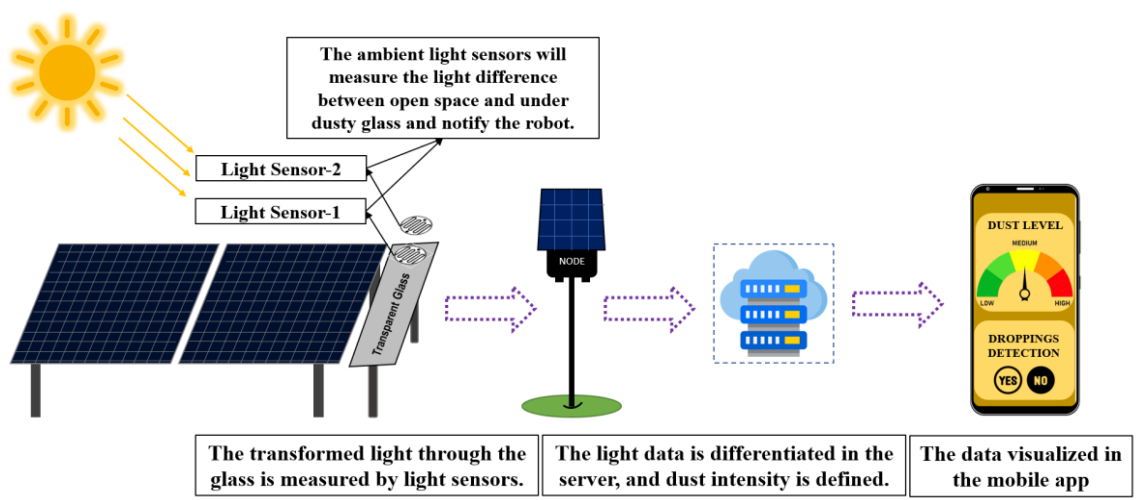

**Fig. 6.** Dust detection system by using light sensors in daylight

Fig. 7 presents a versatile dust detection system that operates using light sensors at night. The system utilizes a light sensor and a relay switch-controlled DC LED. A clean, transparent glass is positioned beside the solar panel for this method. When dust settles on the transparent glass, the artificial light from the LED is unable to penetrate the glass as effectively. A light sensor placed beneath the transparent glass measures the light that passes through the glass. A relay switch controls the LED, and the microcontroller controls the relay. When there is no sunlight, generally at night, the LED is turned ON, the light sensor under the glass measures the artificial light value, and then the LED is turned OFF. The measured value is then compared with the fixed lux value of the LED. If the measured value is less than the LED lux value, that means all light cannot pass through the glass because of dust or any obstacles. For clean glass, the measured value is equal to the LED lux value. So, the dust level on the surface is measured by differentiating the actual lux value of LED and transformed light lux value through the glass. The LED light source is used instead of sunlight because sunlight angle and intensity change over time, causing unstable values sometimes. Both system's principles are the same for measuring dust accumulation, but light sources are different.

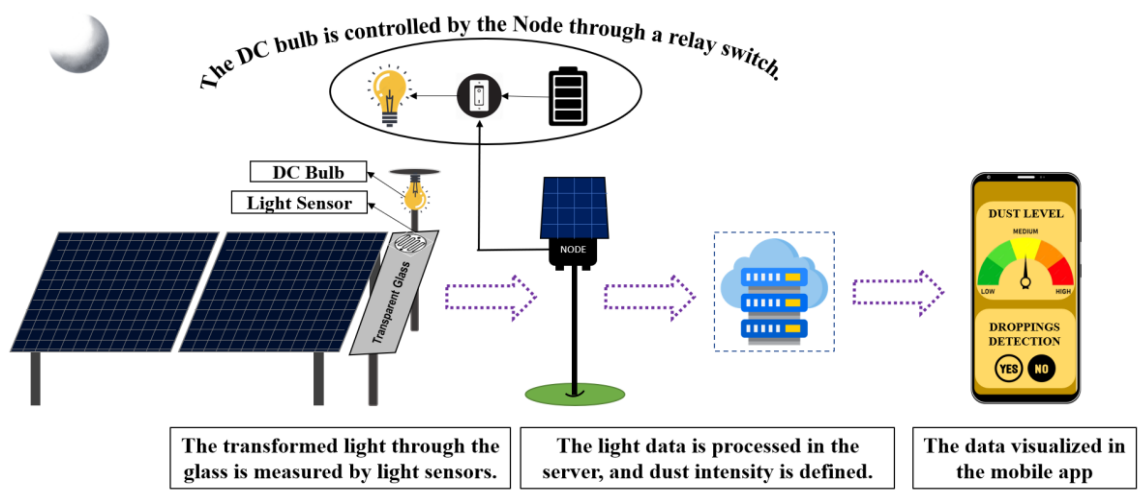

**Fig. 7.** Dust detection system by using light sensors at night

Image processing is another approach to detecting dust. This approach is also used to measure dust amount. Nevertheless, direct solar panel images are not used to detect. A transparent glass acts as a solar panel. Transparent glass is visible from the other side, but if too much dust falls on the glass, it will be invisible from that side. This concept is used here to detect dust. A camera takes images of transparent glass with a white LED light under the glass. The image must be captured at night to resolve the issue of different light conditions. The captured images go through an image process. After image processing, it counts the pixels of images. The clean area remains white after processing, and the dust remains black. The white and black pixels ratio defines how much dust is on the glass surface. A higher number of black pixels means more dust on the surface. The black pixels and white pixels are counted using the image processing method. Fig. 8 shows the dust detection steps through image processing.

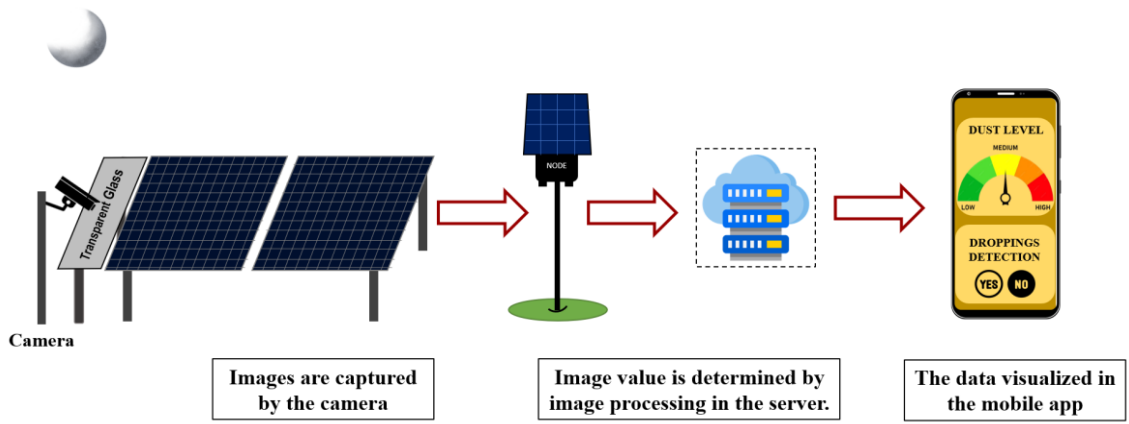

**Fig. 8.** Image processing methods to detect the dust

Fig. 9 shows the complete setup of the dust detection system. In this setup, dust accumulation is measured by a light sensor under transparent glass. The microcontroller of the sensor node then reads the sensor value. After reading the value, it is sent to the cloud. The solar panel images are also captured by the camera and sent to the server for bird-dropping detection by a trained ML model. After processing the light sensor data and getting output for the image in the cloud, the output is sent to the mobile app and visualized.

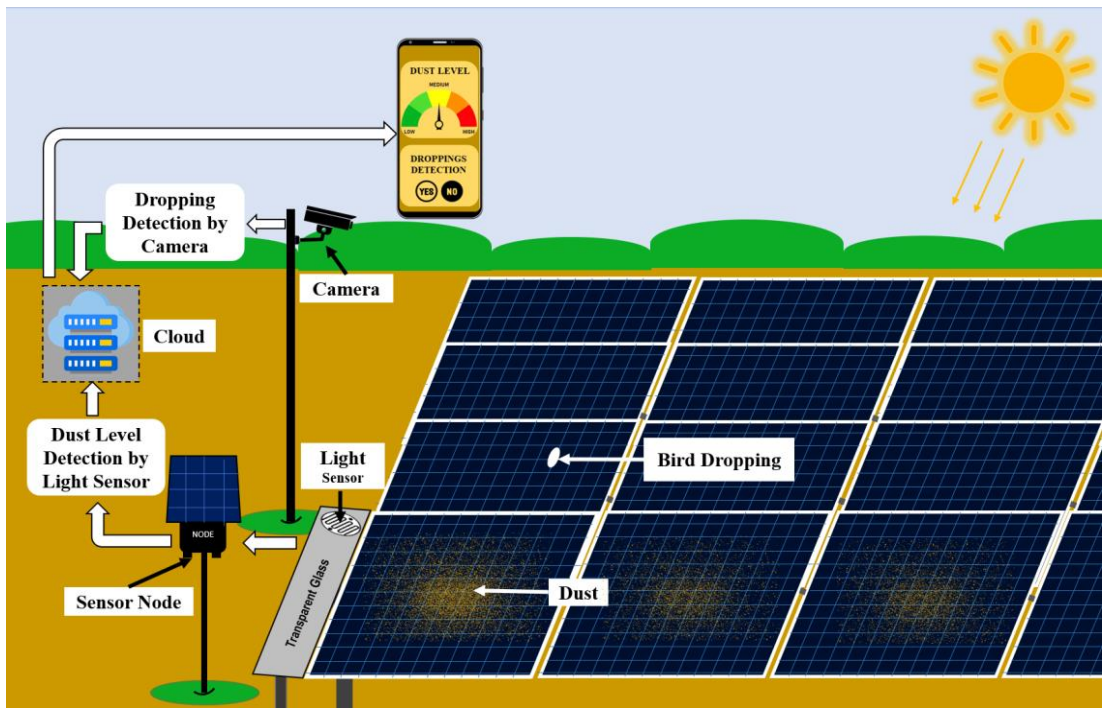

**Fig. 9.** The whole setup of the proposed dust detection method

## *2.3 Field Experiment Setup*

The field experiment setup was arranged at Gazipur (lat. 23.98, long. 90.41). Initially, the setup was for data collection. Solar panels (400W) were positioned in the open field with a

30º tilt angle relative to the horizontal surface. The output power of the panels was taken at a fixed interval before and after cleaning them to evaluate efficiency loss due to dust. A pyranometer was used to get input power to calculate efficiency loss, and data was taken in April. At that time, dust accumulation was also measured by a light sensor to relate and validate the dust accumulation measurement system with efficiency loss. Bird droppings were another common problem faced periodically during the experiment, as seen in Fig. 10a. It also blocked the sunlight, and too much bird waste decreased the output. Fig. 10b, Fig. 10c, and Fig. 10d also show the dust accumulation scenarios.

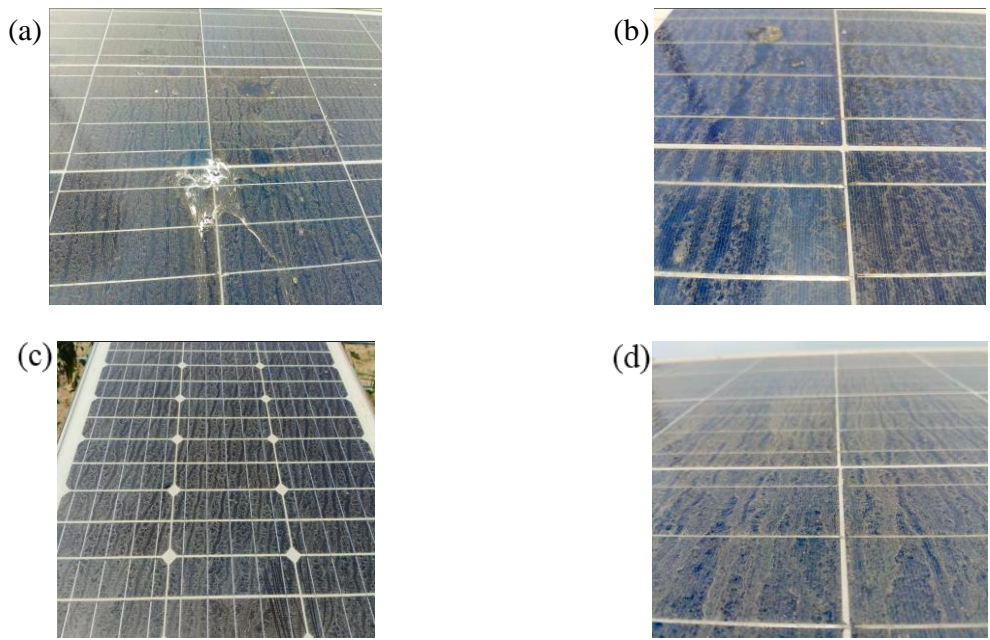

**Fig. 10.** Solar panel covered with (a) bird droppings, (b) insect droppings, (c) soft dust, (d) hard dust

A new approach was used for data taking and validation. Two transparent glasses were placed in the same field. With time, dust falls on the glass and reduces the transparency. Sunlight can pass through clean glass wells but is reflected from dusty glass. The same thing happened for a PV panel. So, the setup was like a PV panel, but instead of solar cells, a lux meter was placed under the transparent glass to measure how much $V_{SL}$ passes through the glass. One glass was regularly cleaned manually to measure the amount of $V_{SL}$ pass-through on the clean glass. Another glass was not cleaned to see how much $V_{SL}$ reflected due to dust. The data for both glasses was taken simultaneously, and both values were compared to get the amount of $V_{SL}$ loss. The loss amount refers to reflected light due to dust. This setup helped to observe the intensity of dust and the falling rate of dust. Initially, a lux meter was used to collect data, and later, two IoT-enabled light sensors were used to get data through IoT. Fig. 11a and Fig. 11b show the dust accumulation on transparent glass, which is measured by a lux meter.

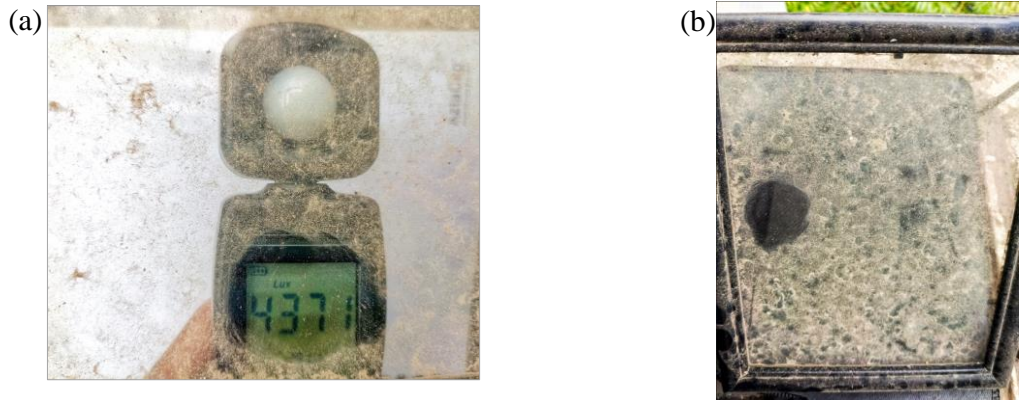

**Fig. 11.** Field setup (a) transparent glass with light sensor, and (b) dust accumulation on the glass

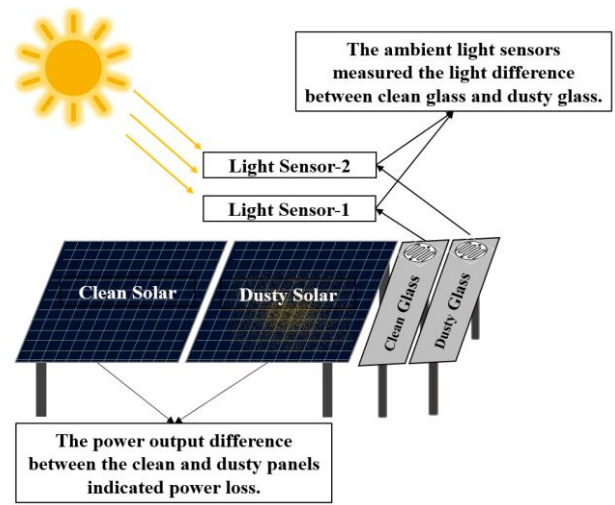

**Fig. 12.** Experimental field setup scenario

## *2.4 Software and ML-based Bird Droppings Detection Implementation*

Data visualization on mobile is essential to get notified about dust accumulation conditions. Sensor data from the field was received and visualized in the mobile app using the Blynk IoT platform. The microcontroller was connected to the Blynk server via template ID and authentication token to send field data. Field data was sent at fixed intervals using the Blynk timer function in the microcontroller firmware. The mobile app was also connected to the machine learning model. Fig. 13 illustrates the step-by-step workflow for getting ML predictions to the mobile app. First, the camera captured and sent the solar panel image to the cloud. The received image was saved in the cloud storage bucket. When a new image was saved in the cloud, a function triggered the load of the pre-trained TensorFlow model. The saved image was processed by resizing, normalizing, or reducing resolution according to model suitability. Then, the image was fed into the model, and the model gave an output. When the droppings were detected, the model gave 1. Otherwise, it gave 0. The result was stored in the cloud. An API was written to fetch the output from the cloud. It was a GET API, and it was inserted in the Blynk webhook. The Blynk console event triggered the webhook for execution. When the function was executed, it received the model output result from the server and posted it to the specific datastream to display in the mobile app.

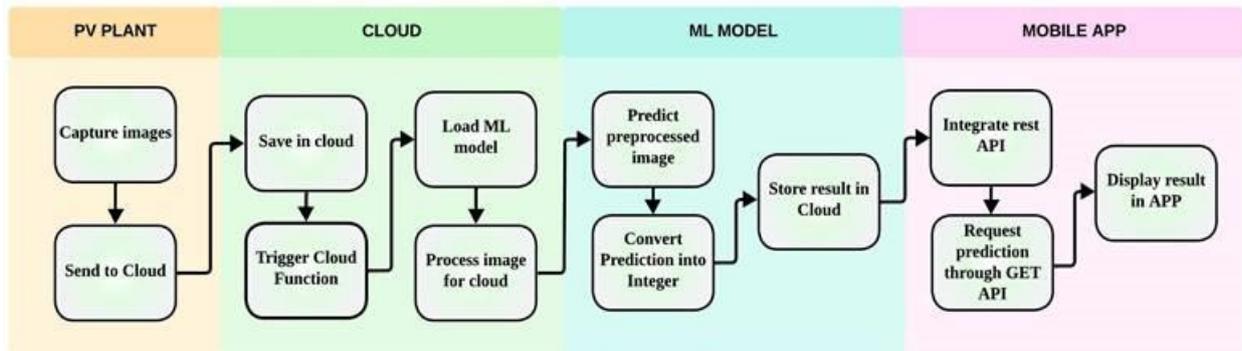

**Fig. 13.** Workflow for ML model prediction

## 3. Results

### 3.1 Application of Mobile App

Fig. 14 shows the data visualization system in the Blynk web dashboard. It has a real-time data visualization system to monitor sensor data remotely, enabling users to monitor PV panel conditions. It also has an app with a real-time notification alert system to notify critical events, and the frequency of the notifications is also customizable. When the dust level exceeds the range, it sends continuous notifications to the user to take action. The implementation of an IoT-based dust detection system enables users to assess the cleanliness of PV panels quickly.

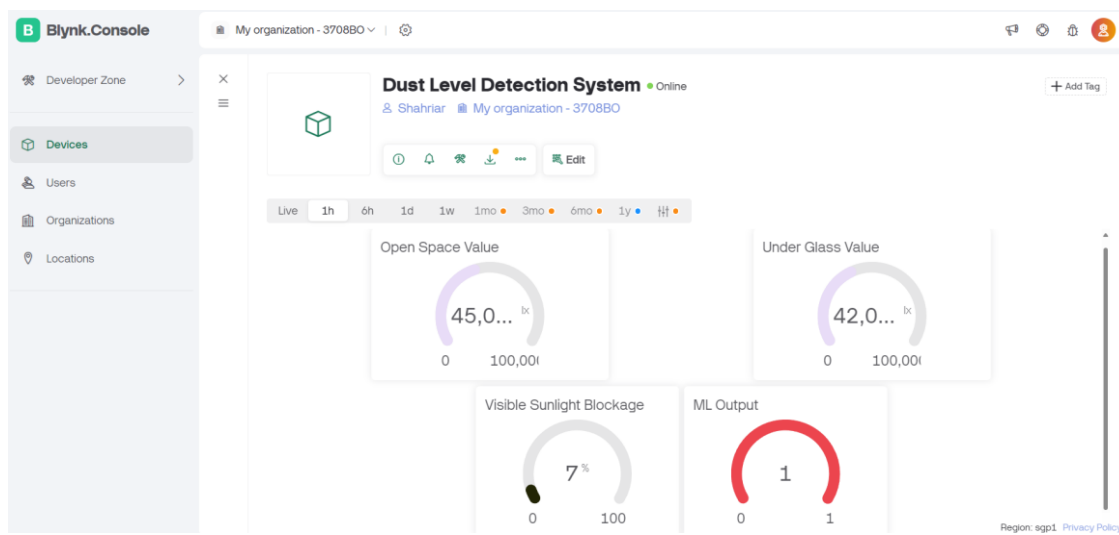

**Fig. 14.** Data visualization in Blynk Dashboard

## 3.2 Dust Level Measurement by Light Sensor

Initially, the team collected the field sunlight data through a handheld lux meter. The data started to be taken from the eighth installation day after some dust accumulation, once daily at fixed times. Fig. 15 shows two types of data in two bars. The blue bar defines the lux value of $V_{SL}$ passes through the clean glass. The red bar represents how much $V_{SL}$ passes through dusty glass or not clean glass. From the first reading, the dusty glass lux value started to decrease. For different sunlight conditions, both values increased or decreased and correlated. The $V_{SL}$ transfer amount for both glasses is differentiated by a relative change formula ($\frac{2nd\ value - 1st\ value}{1st\ value}$) and expressed in percentage. The percentage increased with the day count because more dust regularly fell on the glass panel. More dust obstructs the $V_{SL}$ from passing through the glass. The second graph of Fig. 15 also shows how much the visible light-blocking percentage is increasing with time. Initially, the rate was around 2%; after 20 days, it became 25%, and after 33 days, the value increased to 34%. Here, 2% of $V_{SL}$ blockage means low dust accumulation, and 34% of $V_{SL}$ blockage means high dust accumulation. So, high dust accumulated and blocked 34% of the total $V_{SL}$ within one month (March).

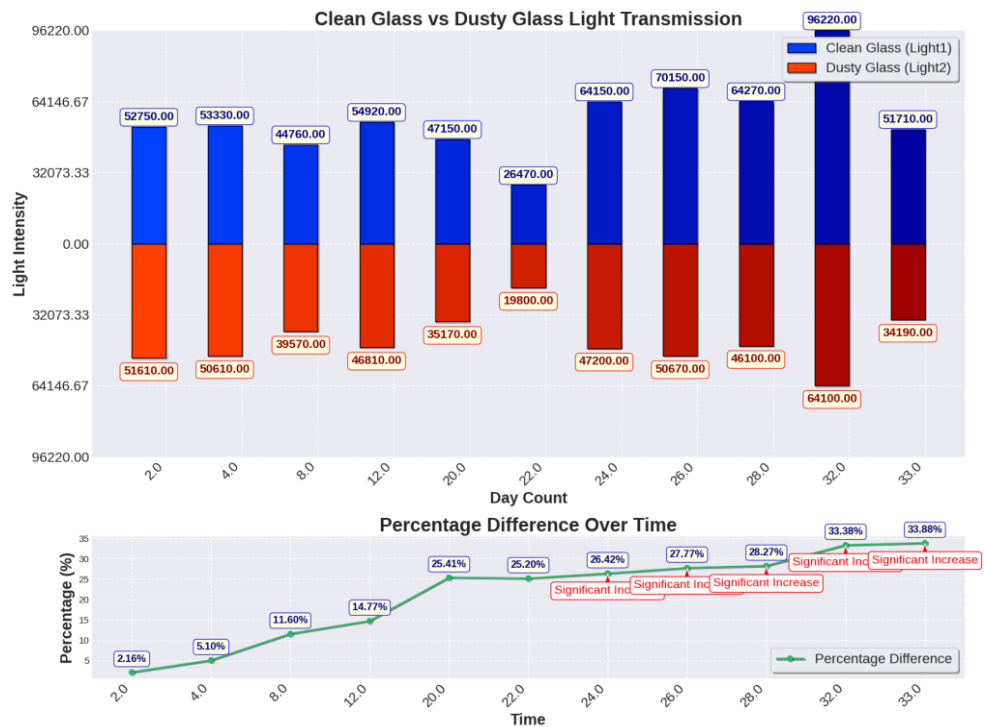

**Fig. 15.** The data of visible sunlight that passes through clean glass vs dusty glass within one month

The team took the second phase of light data, and the approach differed. Fig. 16 shows lux meter data for clean and dusty glass panels. The blue bar shows the light data for clean glass, and the green bar shows the data for dusty glass. The X-axis label defines the day count from the setup. After one month (April) of a certain amount of dust falling on one glass, the data was taken at a 2-3-day interval in May. This approach follows to see the highest $V_{SL}$-blocking percentage for heavy dust. As some dust fell already, the relative change percentage was high from the first data. Fig. 16 percentage graph shows how the $V_{SL}$ blocking percentage is increasing with time. Initially, the sunlight-blocking percentage was 34.59%. After 45 days, the $V_{SL}$ obstructing amount reached 41%. In the next 15 days, after heavy dust fell on the glass, it blocked above 55% of $V_{SL}$. So, it indicates more $V_{SL}$ blocking due to heavy dust fall, which can cause a vast efficiency drop in the PV-based system.

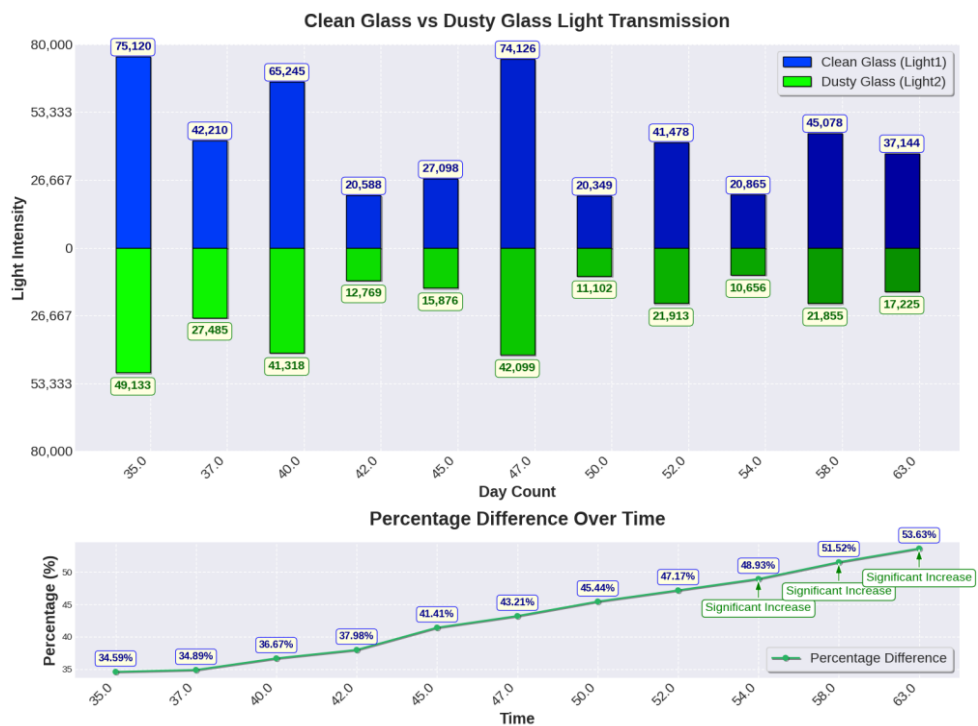

**Fig. 16.** The data of sunlight that passes through clean glass vs dusty glass after one month

*3.3 IoT Light Sensor Data for Different Levels of Dust*

This time, two ambient light sensors were used to automatically collect light data for clean and dusty glass through IoT, and the light loss was calculated. Fig. 17 shows the visual light transmission data for clean and dusty glass for a whole day. The dusty glass had not been cleaned for one week (April). The graph is plotted to see how much $V_{SL}$ varied throughout the day and show a single-day loss. According to the output, clean glass (yellow) has a higher $V_{SL}$ transmission throughout the day. The graph has also demonstrated that, for low light intensity conditions, the loss of $V_{SL}$ is relatively low in dusty glass, even though the same amount of dust was present. However, the loss is comparatively high for the high intensity of the Sun. The loss reached 9.86% at midday but fell to under 9% when the cloud arrived or in the afternoon. This happens because the Sun's position changes throughout the day, so the incoming light angle changes. When sunlight strikes a glass surface, the angle at which it hits (the angle of incidence) influences how much light is reflected versus how much is transmitted through the glass. The graph shows less loss in the afternoon, about 6.17% for low-intensity light. On average, 8.44% of $V_{SL}$ was obstructed throughout the day after not being cleaned for a week.

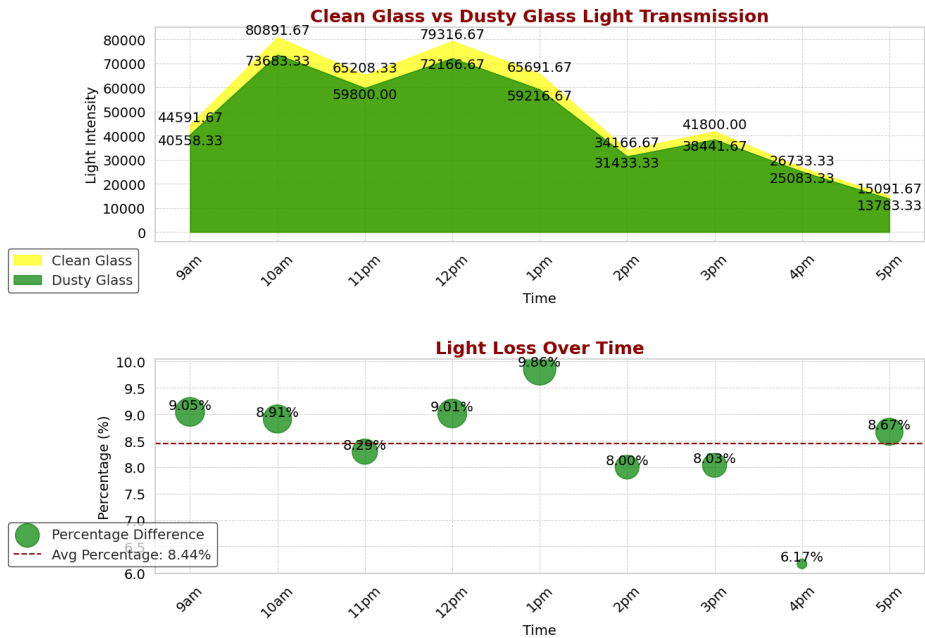

**Fig. 17.** A full-day scenario of sunlight blockage percentage for one week of not-cleaned glass

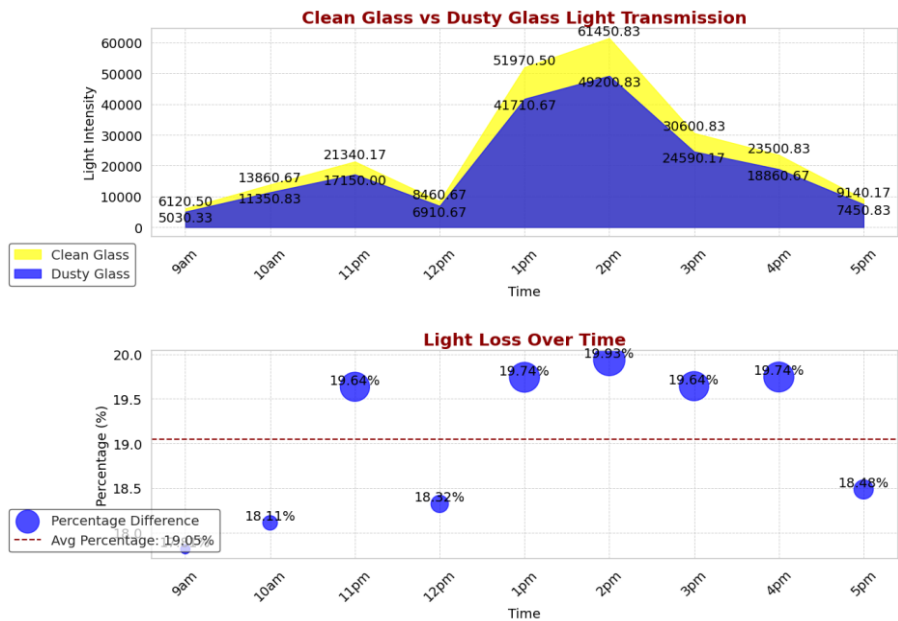

**Fig. 18.** A full-day scenario of sunlight blockage percentage for fifteen days of not-cleaned glass

Fig. 18 highlights the data of dusty glass that was not cleaned for fifteen days. Here, the loss amount (clean glass vs. dusty glass light transmission) is also not constant all day. The loss is high at midday and decreases at the end of the day. The light blockage or loss percentage was above 19% at midday and decreased to 18.48% at evening. So, on average, 19.05% of total $V_{SL}$ was blocked throughout the day due to fifteen days of dust falling.

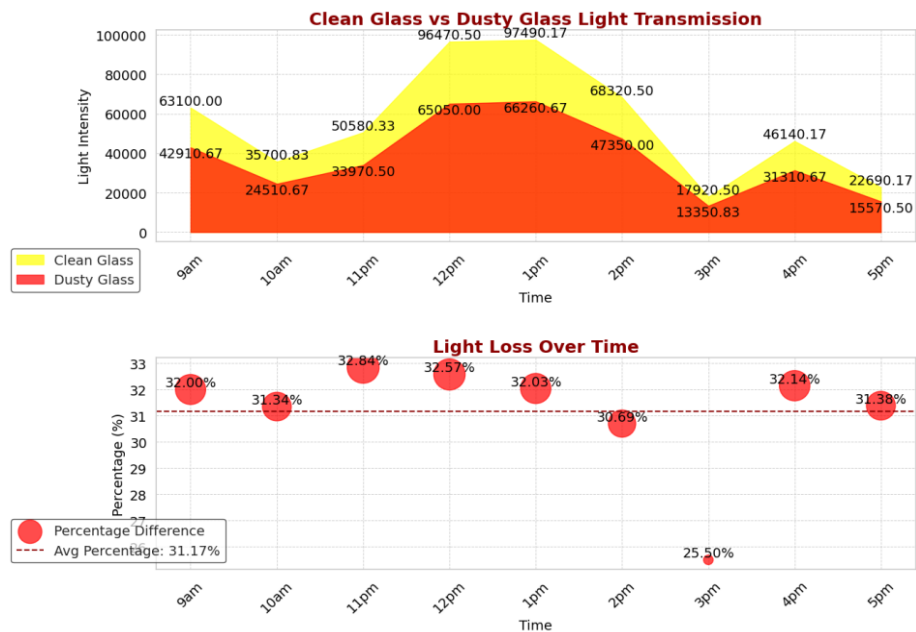

**Fig. 19.** A full-day scenario of sunlight blockage percentage for one month of not-cleaned glass

Fig. 19 represents the data of dusty glass that was not cleaned for one month. After one month of not being cleaned, a high amount of dust fell on the surface, and the light blockage percentage became high. The loss at midday was above 32% and decreased to 25.50% when total light from the sun decreased due to clouds. So, on a single-day average, 31% of total sunlight was obstructed due to dust. The value varies with the daily weather pattern.

The light sensor method defines the first six-month dust accumulation rate, and the data are visualized in Fig. 20. The graph shows that in January, the highest 35% of $V_{SL}$ was blocked after dust accumulated throughout the month. The dust accumulation rate was at its maximum this month. In the next three months, the $V_{SL}$ obstruct percentage was 33%, 33%, and 31%, respectively. So, dust accumulation was reduced a bit compared to January month data. The dust accumulation reduced hugely in May; this time, the percentage was 21%. The weather changing was responsible for that. Since May, the weather pattern has been good due to the rainy season [42]. June had the lowest dust accumulation, which caused only 18% $V_{SL}$ blockage.

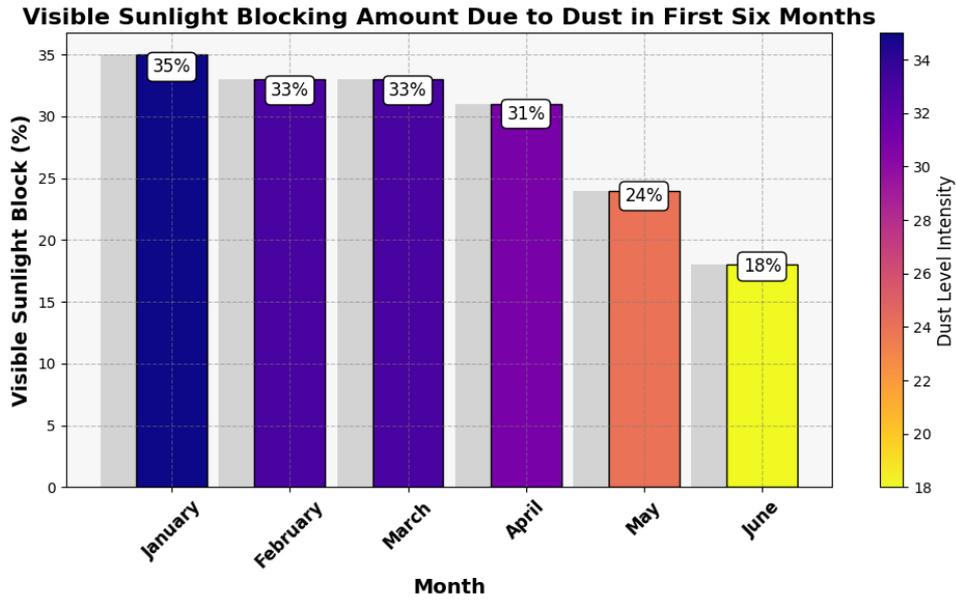

**Fig. 20.** The first six-month scenario of dust accumulation

### 3.4 Dust Effects on PV Panel Efficiency

Solar panel efficiency data were collected from 400W solar panel setups every four days. At the same time, the $V_{SL}$ blocking amount was measured to see if its value changed with the solar panel efficiency. Data on apparent sunlight blocking and solar panel efficiency loss in April are shown in Table 1.

**Table 1.** Solar panel efficiency loss with visible sunlight obstruction data due to dust in April

| Date | Visible sunlight blockage (%) | *Solar panel efficiency loss (%) |
|---|---|---|
| 5/4/2024 | 3.86 | 2.06 |
| 9/4/2024 | 3.8 | 2.09 |
| 13/4/2024 | 4.1 | 2.13 |
| 17/4/2024 | 5.05 | 3.30 |
| 21/4/2024 | 3.08 | 1.83 |
| 25/4/2024 | 4.45 | 2.24 |
| 29/4/2024 | 4.6 | 2.28 |
| 3/5/2024 | 3.2 | 1.96 |

*Solar panel efficiency was calculated by considering Pyranometer solar irradiance value as input power and solar panel output as output power. The efficiency loss was calculated by comparing the solar panel output data before cleaning and after cleaning. As efficiency recovered after cleaning, it lost efficiency due to dust.

More significant solar panel efficiency loss is seen when sunlight blocking rises. For example, on April 17th, the enormous $V_{SL}$ blockage was 5.05%, followed by the most significant efficiency loss of 3.30%. The efficiency loss was reduced to 1.83%, with a low quantity of $V_{SL}$ obstruction of 3.08% on April 21. So, this data indicates that solar panel performance over time corresponds to obstruction of $V_{SL}$ and validates the dust accumulation measurement method.

*3.5 Rainfall Effects on Dust Accumulation*

Fig. 21 shows the June rainfall data and dust-fall scenario on a surface. The whole month scenario summarizes that rainfall helps to clean the PV panel naturally. That means regular rainfall ensures the PV panel's effective efficiency.

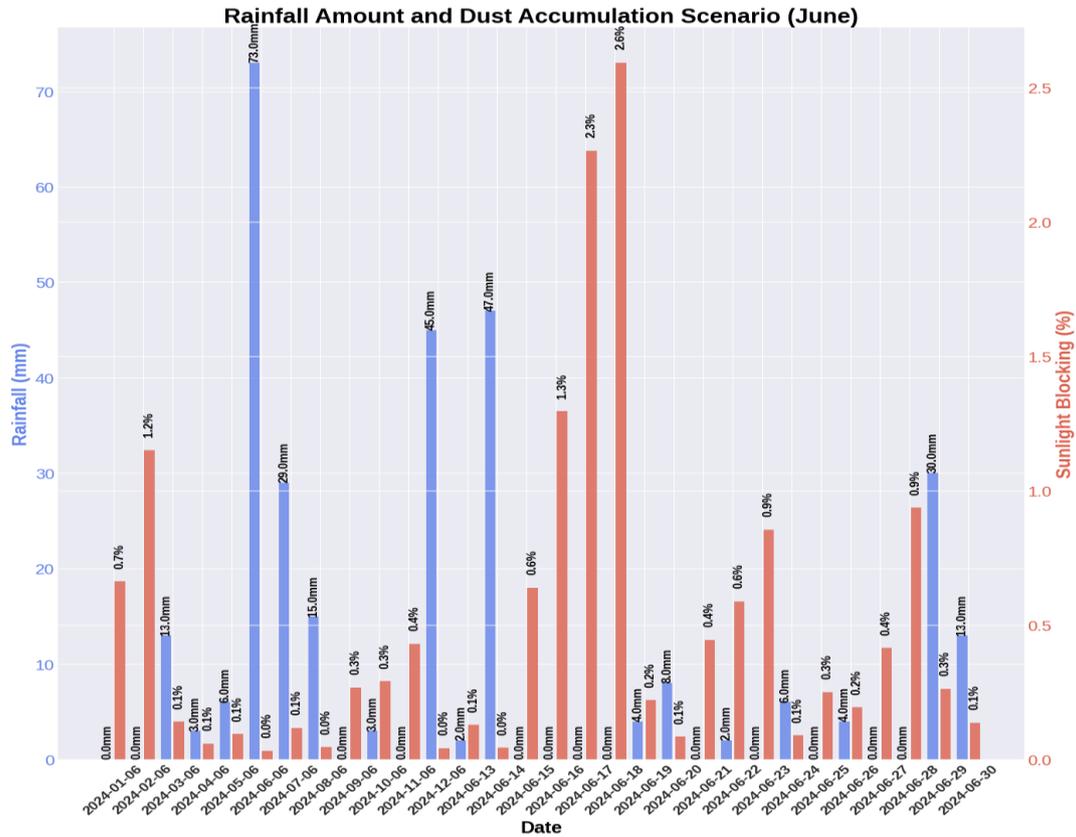

**Fig. 21.** Effects of rainfall on dust accumulation rate

The two glasses and the light sensor took the dust accumulation data. In the first two days of June, no rain occurred. That is why 1.2% of $V_{SL}$ was blocked due to dust accumulation. Nevertheless, continuous rain in the next ten days helped to keep the surface clean. After that, no rain occurred from 15th June to 19th June, and dust accumulated on the surface. Dust prevented about 2.6% of $V_{SL}$ from being on the glass surface. From 20[th] to 28[th] June, rain occurred in small amounts at an interval, but it also helped to clean the glass. Rain occurred in a total of 17 days in June in different amounts, and the 7-day rain amount was above 20mm, which is enough to clean naturally [23]. The total rainfall amount in June was 303mm. This frequency and amount of rain kept the surface clean, which can prevent decreasing PV efficiency.

*3.6 Dust Detection through Image Processing*

For dust detection by image processing, the dust intensity is divided into three levels: no, medium, and heavy dust. The processing methods are the same for all types of images, which are cropping the images, converting them into grayscale, applying a doc scanner-like filter, converting to a binary image using a simple threshold, calculating the total number of black pixel and white pixel and calculating the percentage of both pixels. The Gaussian blur was not used here but can reduce the noise if the image has lots of noise. First, heavy dusty glass images were taken to process. Fig. 22 illustrates the steps of image processing and gives the output the number of black and white pixels. As it is an image of a dusty surface, it has a higher percentage of black area, 25.19%. Fig. 23 shows the result for low dust images, which has a 17.27% black area percentage. The no-dust images have a 0% black area, as shown in Fig. 24. The rest of the area is white pixels, which defines the clean area.

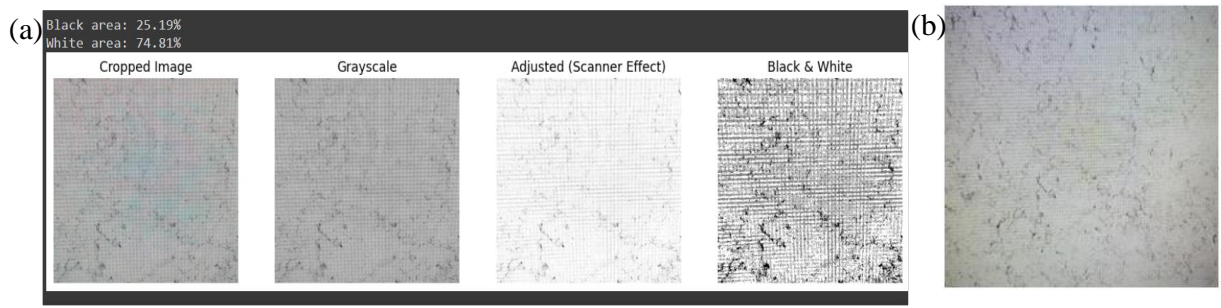

**Fig. 22.** High-level dust detection on the glass surface by (a) image processing and (b) raw image

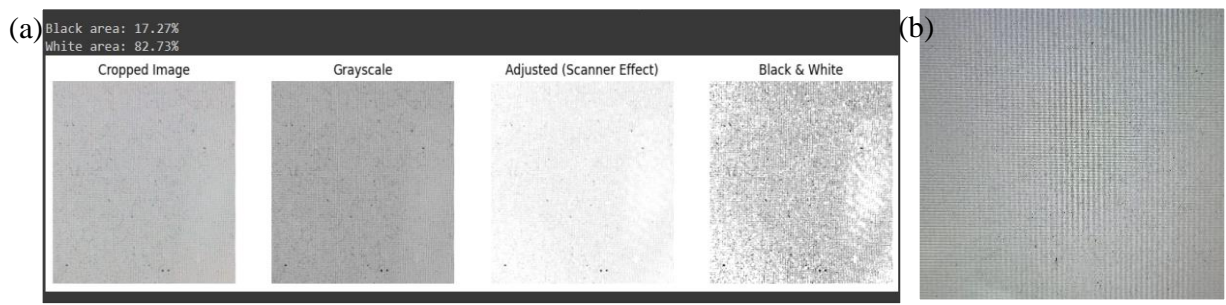

**Fig. 23.** Low-level dust detection on the glass surface by (a) image processing and (b) raw image

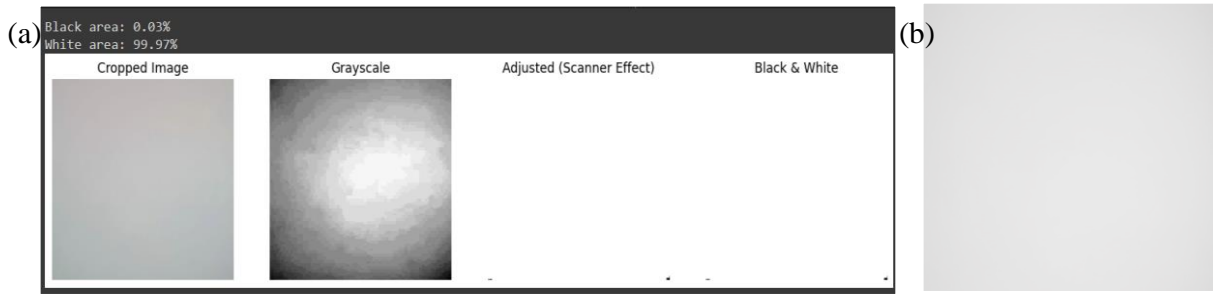

**Fig. 24.** No dust detection on the glass surface by (a) image processing and (b) raw image

*3.7 Bird Droppings Detection through ML Model*

According to Engr. DM Majibor Rahman (Managing Director of Solar Electro Bangladesh LTD), "The droppings of birds and bees are falling on PV panels numerous times, and it happens regularly. It decreases the PV plant's overall output, making it difficult for the workforce to clean. The droppings are sticky and require water to clean them fully" (Doe, personal communication, September 1, 2022). The lab test result also found that the covered area by droppings blocked 70-80% of $V_{SL}$. So, the significance of detecting and cleaning bird droppings is cleared out. An ML-based dropping detection system on PV panels has been implemented to detect them and notify the robot. Tensorflow 2.15.1 was used to build a detection model, an open-source machine learning framework [44]. About 200 images were collected and annotated by labelimg version 1.8.4 to create the custom dataset [45]. The captured images were labeled, and the resolution was converted according to the model requirement using a script in Python version 3.8 [46]. Two hundred images were annotated in PASCAL VOC (Pattern Analysis, Statistical Modelling, and Computational Learning Visual Object Classes) format and saved in a .xml file. There was only one class, which was bird droppings or dust. The EfficientDet D0 512x512 model was used to train. It has good speed (39ms) and good COCO mAP score (33.6) [47]. Before starting the training, some operations were done, like converting XML to CSV, CSV to TF (training file) record, label map formatting, etc. The model was trained on Google Colab using a GPU (Graphical Processing Unit). The computation in GPU was 14x times faster than CPU (Central Processing Unit) for this case. The batch size was set to 16 to reduce loss fluctuation during training. The total step was fixed to 10000, but after 4000 steps, the execution was stopped due to no change in total loss. After completing the model train, the model was tested with test images, and the output is shown in Fig. 25a, Fig. 25b, and Fig. 25c. The model detected all the dust on the panel with 100% confidence. Thus, the model gets a good score in every detection. Fig. 25d is the visual output after calculating the IoU (Intersection over Union) to define the model's performance.

Here, the red-colored box is manually labeled (ground truth), and the yellow-colored box is predicted from the model. This thorough research process instills confidence in the findings, as it ensures that every step is meticulously executed and validated.

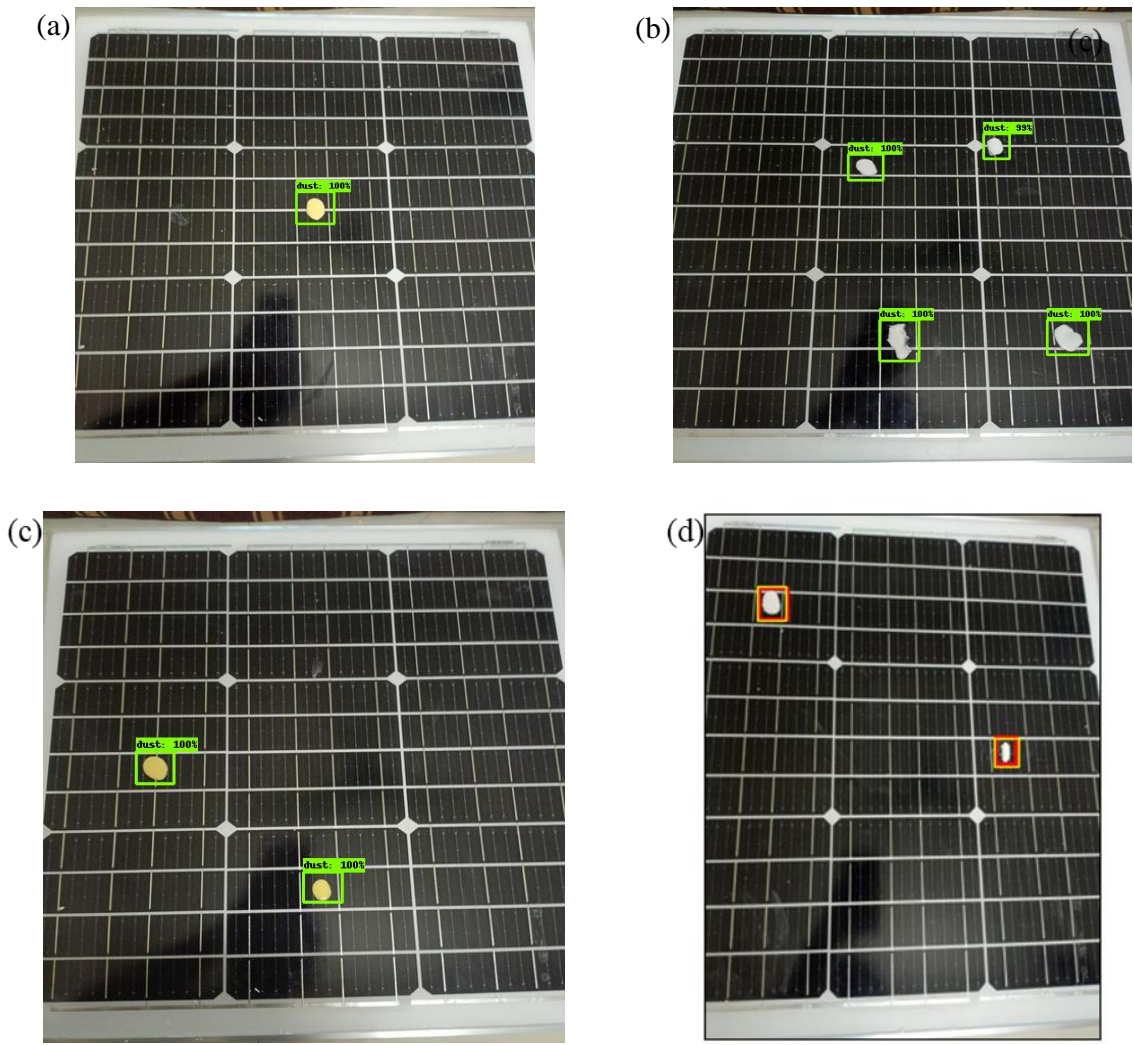

**Fig. 25.** Detection of bird droppings on PV panel (a) single droppings (b) too many droppings on panel (c) different colored droppings (d) IoU calculation for each detected object

The ioU score is a crucial metric in evaluating the performance of the model. It measures the overlap between the predicted and ground truth bounding boxes, with a score ranging from 0 to 1. A threshold is used to determine whether the model successfully predicted the box position. Suppose the model predicts a box with an IoU score equal to or greater than the threshold. In that case, it indicates a significant overlap between the predicted box and one of the ground-truth boxes, signifying that the model detected an object correctly. In the test, the IoU score of two detections was 0.76 and 0.86, both above the threshold of 0.5, as shown in Fig. 26. This means there was a 76% and 86% overlap between the ground truth and predicted bounding boxes, which is a high number. Therefore, our model can successfully detect bird droppings as dust.

```
Intersection over Union (IoU) calculation:
Detection1 (iou, intersect, union): (0.762224242933666, 1675.8000000000004,
  2198.565599999999)
Detection2 (iou, intersect, union): (0.8623840939145838, 2680.7000000000007,
  3108.475699999999)
```

**Fig. 26.** Output of IoU test

**4. Discussion**

Dust accumulation may dramatically reduce efficiency if the PV panel is not cleaned for a long time. Detecting the presence of unwanted objects allows instant cleaning, making the system more efficient. It is also required to automate the cleaning process. A complicated dust detecting system is challenging to construct and hard to run at the raw level. Most AI-based dust detection systems depend on environmental data, which is not always freely available. Many calculations and measurements of solar irradiance are needed to detect dust from power output. This research constructed a dust detection system using only an ambient light sensor under the transparent glass. The light sensor reduced the complexity of dust detection using artificial intelligence. It is less challenging to implement and requires low maintenance. The sensor costs only $5, which has made the system cost-effective. The dust level was indicated by determining the $V_{SL}$ blockage amount. Field data were collected with the light sensor, and it showed that dust blocked 33% of sunlight after not cleaning the surface for one month (in March). Two months of not cleaning caused extensive dust accumulation, blocking up to 55% of the $V_{SL}$. The data for April indicated an 8% $V_{SL}$ transmission reduction after seven days, a 19% reduction after fifteen days, and 31% of $V_{SL}$ blockage after not being cleaned throughout the month. About 303mm of rainfall in 17 days helped to maintain the dust cleaning throughout June naturally and produce sound output. Month-wise $V_{SL}$ blockage measurements have shown that the first four months have a higher dust fall rate. The air quality data of Dhaka also showed that the air quality of the first four months was terrible. Especially in January, the air quality was mostly awful, and this month had the highest dust fall rate. So, the air quality plays a role in the dust accumulation on PV panels. Table 1 validated the dust detection level approach by relating it to the PV panel efficiency. When the $V_{SL}$ obstructed amount was high, efficiency was decreased. When the $V_{SL}$ blockage amount was low, PV efficiency was reduced a bit. Finally, the system could detect bird droppings with higher accuracy.

Solar energy is the most plentiful of all renewable energy sources. Location, weather, tilt angle, shading, temperature, system efficiency, and dust pollution affect solar energy generation in a solar panel installation. Among them, dust pollution can be recovered by cleaning. As the cleaning process is costly in terms of money and water usage, it is more efficient to clean based on sensing rather than routine. This research showed that dust accumulation was much higher in January than in June. So, the frequency of cleaning required more in January than in June. Pollution depends on weather and environmental conditions and changes with them. Therefore, the light sensor-based dust sensing method's dynamic adaptability helps it perform well in such situations.

Dust was detected in two ways: by light sensor and by camera. The light sensor could give more detailed data than the camera because the camera-based system had only limited dust detection levels. However, the light sensor-based detection system is more responsive to different dust levels. By using this, it is possible to analyze daily, weekly, and monthly dust accumulation data. Here, sunlight was used as a light source to measure dust levels. This research found that sunlight angle and intensity change over time, especially in cloudy conditions, which gives an unstable value. However, it was mapped by doing an average. Alternatively, LED sources can also be used to get stable values. The IoT connectivity made it accessible from anywhere worldwide and sent data immediately. The AI-based dust detection system detected bird's and insect's waste with 100% accuracy. Bird droppings amount could not be specified by a few sensors or specific area coverage, which could not give the proper idea. So, a good solution is to cover the entire area with the camera and use an object detection system to detect this dust. An ML model was developed by continuous training with field images for dropping detection and is used here.

There is much scope for future research, especially with AI. The data on air quality taken by the dust sensor and dust accumulation measurement by the light sensor can be utilized for continuous training of the model with other parameters like temperature and humidity. As there is some relationship between dust, temperature, and humidity, the combination will predict the dust fall rate in the future.

**Funding:** This research did not receive any specific grant from funding agencies in the public, commercial, or not-for-profit sectors.

**Data Availability Statement:** The data will be available to the corresponding authors upon reasonable request.

**Acknowledgments:** The authors would like to thank Solar Electro Bangladesh LTD (SEBL) for supporting and allowing us to collect information. Also, thanks to the Bangladesh Agricultural Research Institute (BARI) for allowing us to set up our experiment and instruments in the open field to collect data and evaluate the system's performance.

**Institutional Review Board Statement:** Not applicable.

**Informed Consent Statement:** Not applicable.

**Conflicts of Interest:** The authors affirm that they do not have any conflicts of interest.

**Ethical Statement:** This study did not require ethical approval as it did not involve human or animal subjects.